# An Optimization of Fractal Microstrip Patch Antenna with Partial Ground using Genetic Algorithm Method

Hamid .M.Q. Rasheda[1], Norshahida Mohd Shah[2], Abdu Saif[3], Qazwan abdullah[4], Abbas Uğurenver[5] , Abdul Rashid.O. Mumin[6], Nan Bin Mad Sahar[7]

[1,2,4,6,7], Faculty of Electrical and Electronic Engineering
Universiti Tun Hussein Onn Malaysia
Johor, Malaysia

[3]Department of Electrical Engineering
University of Malaya
Kuala Lumpur, Malaysia

[5] Faculty of Electrical Engineering
İstanbul Aydin University,
Istanbul, Turkey

[1]hamidmohamed23@hotmail.com

*Abstract*: **UWB is increasingly advancing as a high data rate wireless technology after the Federal Communication Commission announced the bandwidth of 7.5 GHz (from 3.1 GHz to 10.6 GHz) for ultra-wideband (UWB) applications. Furthermore, designing a UWB antenna faces more difficulties than designing a narrow band antenna. A suitable UWB antenna should be able to work over the Federal Communication Commission (FCC) of ultra-wide bandwidth allocation. Furthermore, good radiation properties across the entire frequency spectrum are needed. This paper outlines an optimization of fractal square microstrip patch antenna with the partial ground using a genetic algorithm at 3.5 GHz and 6 GHz. The optimized antenna design shows improved results compared to the non-optimized design. This design is optimized using a genetic algorithm and simulated using CST simulation software. The size of the optimized design is reduced by cutting the edges and the center of the patch. The optimized results reported, and concentrated on the rerun loss, VSWR and gain. The results indicate a significant enhancement as is illustrated in Table II. Thus, the optimized design is suitable for S-band and C-band applications.**

I. INTRODUCTION

Due to some advantages of microstrip patch antenna such as lightweight, low profile, cost-effectiveness, it is commonly and widely used [1-2]. However, it's very important nowadays to smallness the antennas because of the vast request for small antennas for many applications such as medical applications, mobile applications and so on. A small antenna is needed to be combined into extremely compact to produce mass quantities handset [3-5]. Besides, due to the rapid development of communication, the microstrip patch antenna is a very special type of antennas and it fulfills further necessities which may make the design steps and process much more convoluted. Hence, the problem of designing the antenna concerned a great number of parameters that have a large result on the behavior of the antenna [6-8].

The fractal structure is the ingredient part of the shape, and the reason for using a fractal shape is to reduce the total dimension of the patch by escalating the efficiency which meets up its requirements volume with electrical length. The proposed fractal is analyzed where the border is close to one wavelength. Also, the fractal shape design not only has a huge effective length as mentioned before but other than the outline of the design also generates an inductance or capacitance which can help to match the designed antenna to the circuit. There are several shapes of a fractal that can be applied or proposed [6], [9-10].

The hexagonal honeycomb shape fractal antenna with DGS was designed, simulated, and analyzed for UWB applications by [11]. The basic design concentrated on hexagonal rings sorted in honeycomb shape along with the partial ground. To enhance the bandwidth, the ground was integrated with a triangular slot and the hexagonal rings were loaded by hexagon patches. The authors in [12] used Differential Algorithm (DE) to design a new inset-fed slotted microstrip patch antenna. DE was employed to improve the slotted patch antenna design parameters according to needs and scenarios, as well as reduce the computational complexity of the antenna. The application of a slot in the ground plane, and Defected Ground Structure (DGS) of a microstrip patch antenna were offered by [13-14]. The goal of this technique is to mitigate the higher-order excited patterns without degradation of the antenna parameters in the basic mode. The authors introduced the method for the designing of the slot in H-format. This method was developed according to the physical dimensions of the microstrip patch antenna. The location of the DGS-related feed line was set based on a detailed parametric analysis. The authors in [15-16] developed two antenna prototypes, with and without DGS. The authors in [16-17] also carried out measurements of reflection coefficient and radiation pattern. This paper designs both a single microstrip patch antenna and a fractal patch antenna. The patch antenna and the fractal shape are designed for the C band, and both designs are modified and optimized to improve the outcome by genetic algorithm and using CST software [18-21]. The size of the optimized design is reduced by cutting the edges and the center of the patch. The remainder of this paper is organized as follows. Section II explains in detail the antenna design. Experimental results are presented and analyzed in Section III, while Section IV concludes the paper.

## II. ANTENNA DESIGN

In this section, the design stages together with the equations of the design are outlined. The parameters of the presented antenna design are calculated at 7 GHz. It is very important to choose well substrate materials for the antenna because the dielectric substrate has a significant impact on the impedance and the radiation pattern of the patch design. Moreover, the loss tangent, dielectrics constant and other electrical properties are considered the utmost essential characteristic when selecting a substrate. The design of the proposed design is inversely proportional to the constant of a substrate. In this paper, FR4 has been utilized for the design of a square microstrip patch antenna with inset feed with an operating frequency at S-band and C band, and a dimension size of $10 \times 10\,mm^2$. The thickness of the substrate is 1.57 mm as indicates in Fig. 1(a) indicates patch and (b) indicates ground plane.

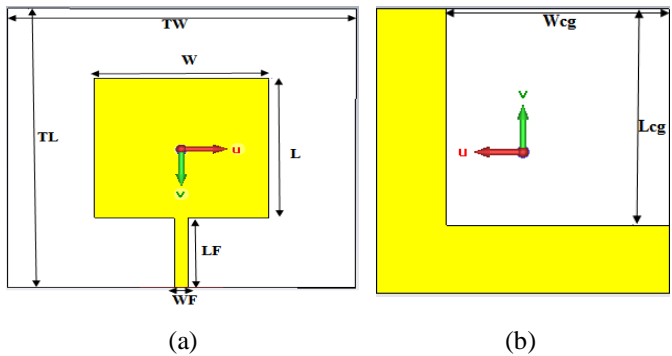

Fig 1: (a) geometry of patch antenna and (b) Ground Plane (unoptimized design)

It determines the diminution of the patch by applying the mathematical equation before start simulation by using the following equations:

a) Calculation of the patch width

$$W = \frac{C}{2f_r \sqrt{\frac{(\varepsilon_r + 1)}{2}}} \quad (1)$$

b) Calculation of effective dielectric constant

$$\varepsilon_{reff} = \frac{\varepsilon_r + 1}{2} + \frac{\varepsilon_r - 1}{2} \left( \frac{1}{\sqrt{1 + 12\left(\frac{h}{w}\right)}} \right) \quad (2)$$

c) Calculation of length extension

$$\Delta L = 0.412 h \left( \frac{(\varepsilon_{reff} + 0.3)\left(\frac{W}{h} + 0.264\right)}{(\varepsilon_{reff} - 0.258)\left(\frac{W}{h} + 0.8\right)} \right) \quad (3)$$

d) Calculation of the actual length

$$L = \frac{c}{2f_r \sqrt{\varepsilon_{reff}}} - 2\Delta L \quad (4)$$

Now after the first design of the square microstrip patch antenna, fractal shape design is done and comes into an illustration. Thus, in this step, the genetic algorithm is used to optimize the proposed square microstrip patch antenna, and the shape is divided into $N \times N$ small square shape elements are taken out and the rest remained the same. So, by optimizing few elements from the patch, the element with "black color" is taken out as shown in Fig. 2 (left side) with W1 and L1 dimensions of 4 mm and 2 mm, respectively, while for W2 and L2 is 2mm × 2mm. Then the optimized design using CST is shown in Fig. 2 (right side). Table I indicates the parameters and their values.

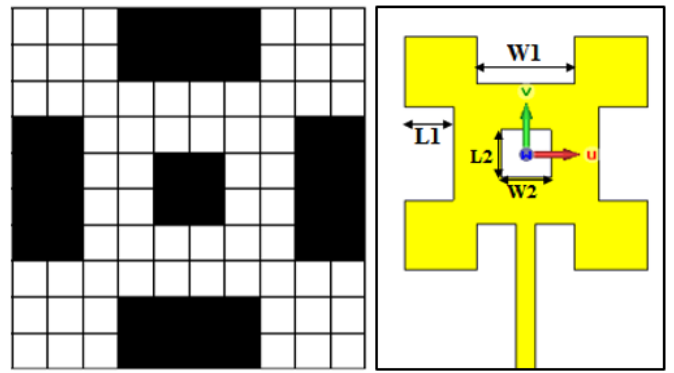

Fig 2: Fractal Geometry (Optimized Design)

TABLE I. SPECIFICATIONS OF THE PROPOSED ANTENNA

| Parameters | Values |
|---|---|
| Length of substrate | 38.7 |
| Width of substrate | 39 |
| Length of the ground plane (reduced), Lg | 15.5 |
| Slot ground plane, Gw | 3.5 |
| Length of a patch antenna, L | 19 |
| Width of a patch antenna, W | 19 |
| Length of Stair 1, Ls1 | 1.5 |
| Width of Stair 1, Ws1 | 14.1 |
| Length of Stair 2, Ls2 | 1 |
| Width of Stair 2, Ws2 | 12.1 |
| Length of the feed, FL | 17.5 |
| Width of the feed, FW | 3.5   mm |

Thus, there are two steps are applied in this design to obtain the optimization using a genetic algorithm which are:
i. Transpose the design parameters into the proper individual. The reason for using a genetic algorithm in this step is to pick the coding of the parameters into genes and binary coding is utilized here. As mentioned before that the patch is divided into $^{N \times N}$ small square elements. Every element is symmetric to a gene of an individual. If the element is taken outset the gene to be "Zero", else set to "One". Hence, the parameter of the geometry design patch is transposed into N by N matric in a binary gene.
ii. Solve model in CST software by setting solving conditions.

III. RESULT AND DISCUSSION

As a result, after finding out the theoretic implementation, it's time to look onward in extracting the performance of few antenna parameters with a help of a domain solver in CST. The genetic algorithm process is completed once achievement the optimization of the parameters.

*A.* **A. Single Patch**

The configuration of the single patch is a square feed by inset fed line as shown in Fig. 1. So, this designed antenna is simulated at 6 GHz and is simulated in educationally available CST Microwave Studio simulation software. The Performance of the antenna design is analyzed using the parameters Return loss, VSWR, and Gain. Return loss is the proportion of the power provide for an antenna to the power returned to the feed step. Consequently, the power provides for an antenna should be engaged rather than being returned, which presents a power loss. Still, the preferable value of return loss shall be negative infinity if the power is taken in and need to be equivalent to zero if the power is returned. Frequently, return loss supports in defining the frequency of any design antenna. Moreover, Fig. 3 indicates the return loss simulated results of -12.8dB at 3.5 GHz and -9.07dB at 6.7 GHz.

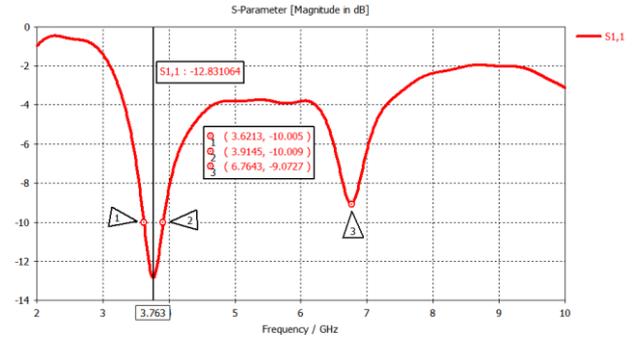

Fig 3.: Return Loss of Unoptimized Design

Figure 4 illustrates the result of VSWR of the unoptimized design which is 1.599 at 3.5 GHz and 2.0865 at 6 GHz. Thus, the VSWR assists in finding the impedance matching of the proposed design with the line of transmission and it should be near to 0 and less than 2.

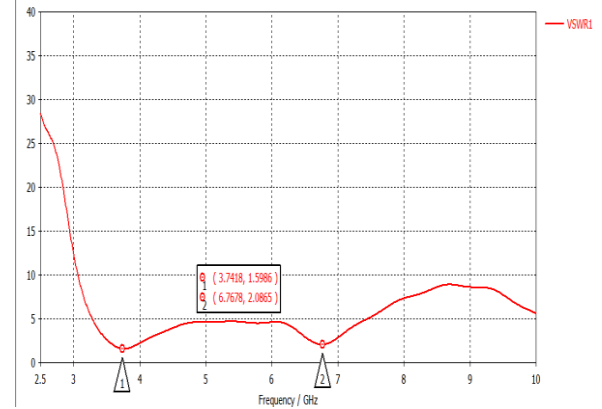

Fig 4: VSWR of Unoptimized Design

Figures 5a and 5b show the simulated gain of the unoptimized antenna, it's clear from the figure that the highest gain attained is 1.52 dB at 3.5 GHz.

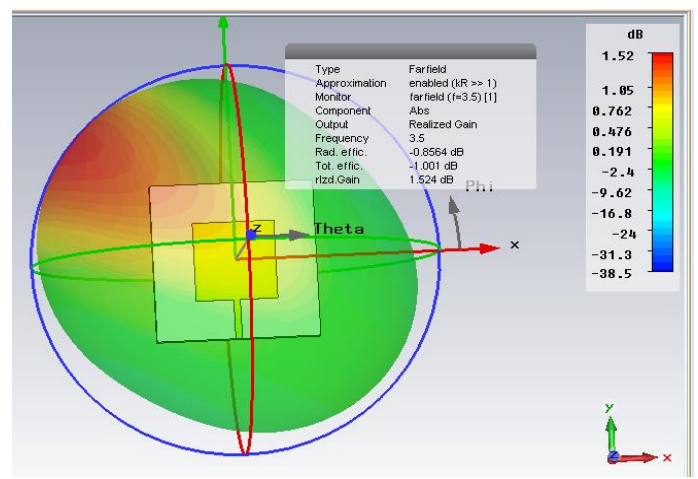

Fig 5: Gain at 3.5 GHz of Unoptimized Design

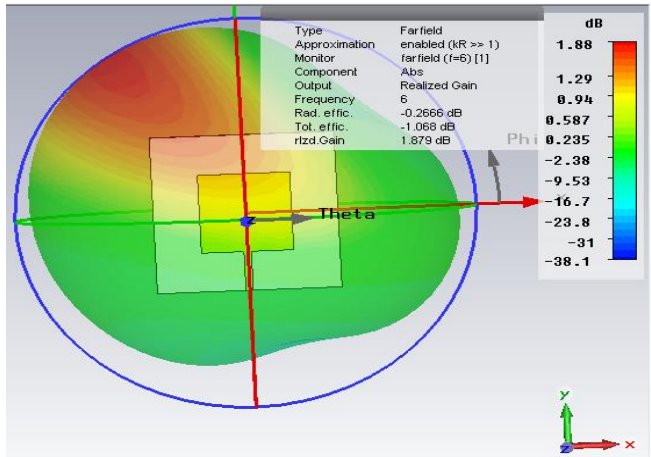

Fig 5:b Gain at 6 GHz of Unoptimized Design

### B. B. Fractal Patch

Figure 2 shows the optimization design (fractal shape) using CST, which is designed at a frequency of 3.5 and 6 GHz and Figure 6 reveals the $S_{11}$ of the proposed design. Figure 7 illustrates the return loss of optimized design which operated at dual-band. At 3.5 GHz the return loss is -12.968 dB, and -12.184 at 6.1 GHz.

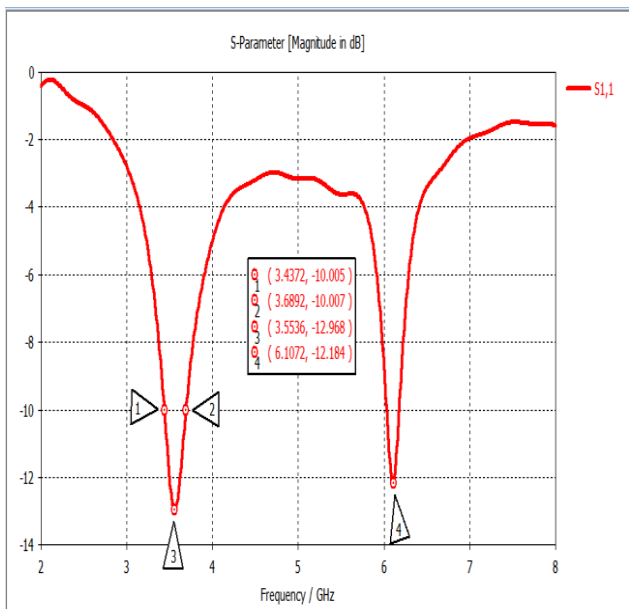

Fig 6. Return Loss of optimized Design

A VSWR value less than 2 is considered appropriate for most antenna applications. The antenna can be described as a "Good match.

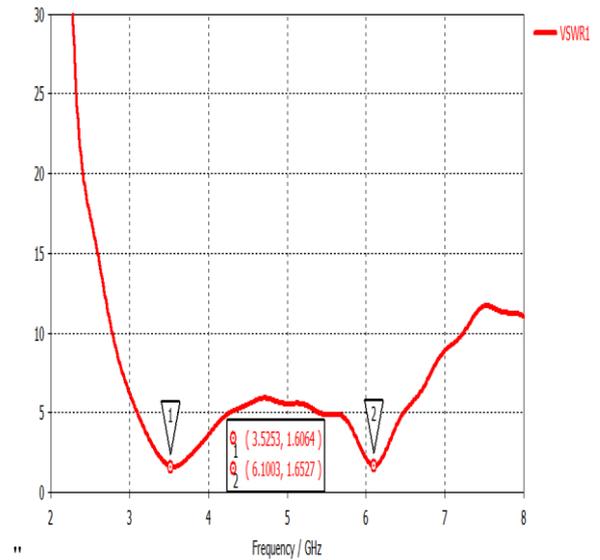

Fig 7. VSWR of optimized Design

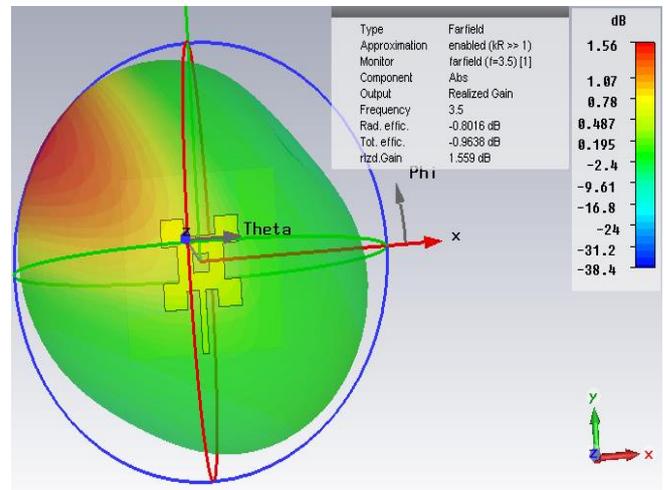

Fig 8a. Gain at 3.5 GHz of Optimized Design

Figures 8a and 8b show the 3D simulated gain of optimized design. It's very obvious that the gain is 1.56 dB for 3.5 GHz, and 2.2 dB at 6 GHz. Table 2 presents the comparison between the optimized and non-optimized designs at a dual-frequency which are 3.5 GHz and 6.1 GHz. The size of the optimized antenna design is reduced and optimized the performance characteristics. The results of the optimized design show the return loss is -12.968 dB and -12.184 dB at 3.5 GHz and 6.1 GHz, respectively, compared to the unoptimized design which is -12dB at 3.5GHz and -9.08 dB at 6 GHz with an improvement of 8.067% and 34. 185% correspondingly. Similarly, 0.313% and 26.195% improvement are very obvious if we compare with the VSWR parameters. A gain is improved by 2.632% and

17.021%. Hence, the optimized characteristics make the optimized design proper for S-band and C-band.

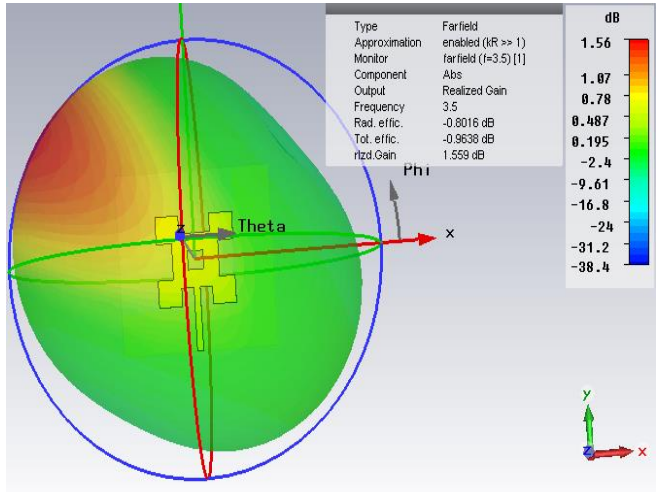

Fig 8a: Gain at 3.5 GHz of Optimized Design

TABLE II: COMPARISON RESULTS BETWEEN UNOPTIMIZED AND OPTIMIZED DESIGN

| Parameter | Unoptimized Design at 3.5 GHz and 6 GHz | Optimized Design at 3.5 GHz and 6 GHz | Percentage Increase in Optimized Design at 3.5 GHz and 6 GHz |
|---|---|---|---|
| Size | $10 \times 10$ mm$^2$ | $6 \times 6$ mm$^2$ | |
| Return loss (dB) | -12. dB and -9.08 dB | -12.968 dB and -12.184 dB | 8.067 % and 34.185 % |
| VSWR | 1.599 and 2.086 | 1.604 and 1.653 | 0.313 % and 26.195% |
| Gain (dB) | 1.52 dB and 1.88 dB | 1.56 dB and 2.2 dB | 2.632 % and 17.021 % |

TABLE III: a comparison with previous

| Reference | Size (mm) |
|---|---|
| 21 | 25 x 25 |
| 22 | 40 x40 |
| 23 | 58 x 52 |
| Proposed structure | 6 x 6 |

## II. CONCLUSION

This paper presents an optimized fractal square microstrip patch antenna with the partial ground using a genetic algorithm at 3.5 GHz and 6 GHz. The size of the optimized antenna is reduced by extracting 4 mm from each side, and 2 mm from the center of the patch. It shows an improvement in the return loss, VSWR, and gain as well. Hence, the optimized design makes the antenna suitable for C-band and S-band application that operates at 3.5 GHz and 6 GHz.

ACKNOWLEDGMENT

This work is supported by the University of Tun Hussein Onn Malaysia under Multidisciplinary Research (MDR) grant vot H470.